\documentclass[preprint,superscriptaddress,amsmath,amssymb,aip,jcp,floatfix]{revtex4-1}
\usepackage[utf8]{inputenc}
\usepackage{graphicx}
\usepackage{enumitem}
\usepackage{gensymb}
\usepackage{amsmath}
\usepackage{amsthm}
\usepackage{amssymb}
\usepackage{dcolumn}
\usepackage{color}
\usepackage{bbm}

\usepackage[normalem]{ulem}

\begin{document}

\title{General analytical nuclear force and molecular potential energy surface from full configuration interaction quantum Monte Carlo}


\author{Tonghuan Jiang}
\affiliation{School of Physics, Peking University, Beijing 100871, P. R. China}

\author{Wei Fang}
\affiliation{State Key Laboratory of Molecular Reaction Dynamics and Center for Theoretical Computational Chemistry, Dalian Institute of Chemical Physics, Chinese Academy of Sciences, Dalian 116023, P. R. China.}
\affiliation{Department of Chemistry, Fudan University, Shanghai 200438, P. R. China}

\author{Ali Alavi}
\affiliation{Max Planck Institute for Solid State Research, Heisenbergstrasse 1, 70569 Stuttgart, Germany}
\affiliation{University of Cambridge, Lensfield Road, Cambridge CB2 1EW, United Kingdom}

\author{Ji Chen}
\email{ji.chen@pku.edu.cn}
\affiliation{School of Physics, Peking University, Beijing 100871, P. R. China}
\affiliation{
 Collaborative Innovation Center of Quantum Matter, Beijing 100871, P. R. China
}
\affiliation{Interdisciplinary Institute of Light-Element Quantum Materials and Research Center for Light-Element Advanced Materials,Peking University, Beijing 100871, P. R. China
}
\affiliation{Frontiers
Science Center for Nano-Optoelectronics, Peking University, Beijing 100871, P. R. China}

\date{\today}

\begin{abstract}

Full configuration interaction quantum Monte Carlo (FCIQMC) is a state-of-the-art stochastic electronic structure method, providing a methodology to compute FCI-level state energies of molecular systems within a quantum chemical basis.
However, especially to probe {\em dynamics} at the FCIQMC level, it is necessary to devise more efficient schemes to produce nuclear forces and potential energy surfaces (PES) from FCIQMC.
In this work, we derive the general formula for nuclear force from FCIQMC, and clarify different contributions of the total force.
This method to obtain FCIQMC forces eliminates previous restrictions, and can be used with frozen core approximation and free selection of orbitals, making it promising for more efficient nuclear force calculations.
After numerical check of this procedure on the binding curve of N$_2$ molecule, we use the FCIQMC energy and force to obtain the full-dimensional ground state PES of water molecule via Gaussian processes regression.
The new water FCIQMC PES can be used as the basis for H$_2$O ground state nuclear dynamics, structure optimization, and rotation-vibrational spectrum calculation. 

\end{abstract}

\maketitle

\section{Introduction}

Calculation of the {\em electronic state} of a molecule, and how it is coupled to {\em nuclear motion}, are two indispensable parts in quantum chemical calculations. 
The potential energy surface (PES) in the Born-Oppenheimer approximation, as a function of the nuclear coordinates, serves as a bridge between the two, and allows further exploration of molecular properties 
such as stable structure, transition path, vibrational spectrum, and nuclear dynamics \cite{martin_2004}.
The PES is also a good starting point to explore effects beyond the Born-Oppenheimer approximation such as electron-phonon coupling, nuclear quantum effects, and non-adiabatic effects \cite{giustino_electron-phonon_2017,sand_analytic_2018,markland_nuclear_2018,nelson_non-adiabatic_2020}.
Therefore, the quality of PES matters and obtaining accurate PES from \textit{ab initio} is fundamentally important.
In standard deterministic approaches, such as density functional theory, the application of Hellmann-Feynman theorem significantly simplifies the problem, making the nuclear force calculation straightforward and efficient \cite{martin_2004}. 
Therefore, these methods can produce both energies and forces on a large scale, so that the PES can be reconstructed, in particular thanks to the latest supports lent by machine learning techniques \cite{unke_machine_2021}. 
However, with more accurate correlated wavefunction methods especially those employing stochastic samplings, the force calculation and PES reconstruction often requires additional treatment besides the already expensive total energy calculation \cite{assaraf_zero-variance_2003,van_rhijn_energy_2022}.

Full configuration interaction quantum Monte Carlo (FCIQMC) is such an state-of-the-art stochastic \textit{ab initio} electronic structure method. \cite{Booth_FCIQMC_2009,Kai_NECI_2020}
With the help of efficient stochastic sampling, FCIQMC solves large full configuration interaction (FCI) type and complete active space configuration interaction (CASCI) problems in much larger Hilbert spaces than the deterministic solvers can handle. 
In the past decade, FCIQMC has witnessed a rapid growth of its own \cite{Cleland_iFCIQMC_2010,Blunt_semi_2015,luo_combining_2018,dobrautz_efficient_2019,Ghanem_adapshift_2019} and inspired the development of other efficient wavefunction methods \cite{tubman_deterministic_2016,holmes_heat-bath_2016,deustua_communication_2018,filip_multireference_2019}.  
Properties such as the one-particle and two-particle reduced density matrices (1-RDM and 2-RDM) can also be evaluated within FCIQMC \cite{blunt_density_2017}. 
Unbiased sampling of RDMs enables evaluation of observables that do not commute with Hamiltonian, such as dipole moments and dipole polarizabilities \cite{Force_FCIQMC_2015}.
Thomas et al. also showed that analytic nuclear forces can be calculated from all-electron FCIQMC calculations\cite{Force_FCIQMC_2015}. 
%
However, restrictions of high computational cost hinder further attempts on obtaining and exploring PES at FCIQMC level. 

In this work, we present a generalization of the FCIQMC force calculation, including frozen orbitals, and apply it to construct PES of water molecule.
In section II we present the formula for FCIQMC nuclear force based on the well-established CASCI energy gradient formula\cite{RHF_CIgrad_Rice_1986,Yamaguchi_1994}, which is applicable to general stochastic CASCI calculations. 
We also formulate the FCIQMC nuclear force formula with restricted open-shell Hartree-Fock (ROHF) orbitals, which has not been shown in literature. 
Computational details of FCIQMC simulation and Gaussian processes regression (GPR), the fitting procedure to obtain PES, are presented in section III.
In section IV, we discuss the results, which contain $\text{N}_2$ force curve and the force map of $\text{H}_2\text{O}$, as well as its full-dimensional potential energy surface. 
%

\section{General analytic force in FCIQMC}\label{theory}

First, we briefly discuss the theoretical background and formulations.
Our theoretical derivation follows and supplements ref. \onlinecite{Yamaguchi_1994} where the energy gradients for Hartree-Fock (HF), configuration interaction (CI) and multi-configuration self-consistent field (MCSCF) have been partially discussed.
Here we present the main formulations and key points. 
More details are discussed in the supplementary information.

In FCIQMC, one presents a stochastic FCI solution to \textit{ab initio} Hamiltonian (Eq. \ref{ab_initio_Hamil}). 
\begin{equation}
    \hat{H} = E_\text{nuc} + \sum_{pq}{h_{pq}\hat{p}^\dagger \hat{q}} + \frac{1}{2}\sum_{pqrm}{(pq|rm)\hat{p}^\dagger \hat{r}^\dagger \hat{m}\hat{q}}
    \label{ab_initio_Hamil}
\end{equation}
where $E_\text{nuc}$ denotes the nuclear repulsion term, $p,q,r,m$ denote molecular orbital (MO) indices, and $h_{pq}$ and $(pq|rm)$ denote 1-body and 2-body integrals, respectively.
Efficient stochastic algorithm allows more efficient solution of large FCI problems, but also adds further complexity to analytic formulations which would apply in deterministic FCI.

%
%
%
%

In general, one can always consider a complete active space (CAS) problem where
several active orbitals and electrons are selected and correlated from the entire 1-electron orbital space.
In this way the 1-electron orbital space is divided into 3 parts: core ($c$), active ($a$) and virtual ($v$) space. 
FCI diagonalization can be performed within the CAS space to generate the CASCI wave function.
The solution approaches the FCI result when the CAS space increases to the full space. 
The total energy of CASCI wave function is formulated as follows. 
\begin{equation}
    E_\text{tot} = E_\text{nuc} + \sum_{pq}{\gamma_{pq}h_{pq}} + \frac{1}{2}\sum_{pqrm}{\gamma_{pqrm}(pq|rm)}
    \label{total_ene}
\end{equation}
where $\gamma_{pq}$ and $\gamma_{pqrm}$ represent the one and two-particle reduced density matrices (1-RDM, 2-RDM), respectively. 
Then the energy gradient of CASCI wave function is
\begin{equation}
\begin{aligned}
    -F_x = \frac{\partial E_\text{tot}}{\partial x} & = \frac{\partial E_\text{nuc}}{\partial x} + \sum_{pq}{\gamma_{pq}h_{pq}^x} + \frac{1}{2}\sum_{pqrm}{\gamma_{pqrm}(pq|rm)^x} \\
    & - \frac{1}{2}\sum_{pq}{S_{pq}^x(I_{pq}+I_{qp}^*)} + \frac{1}{2}\sum_{(p,q)\notin\text{gray}}{(I_{pq}-I_{qp}^*)(U_{pq}^x-U_{qp}^{x*})}
    \label{force_3}
\end{aligned}
\end{equation}
\begin{figure}
    \centering
    \includegraphics[width=9.0cm]{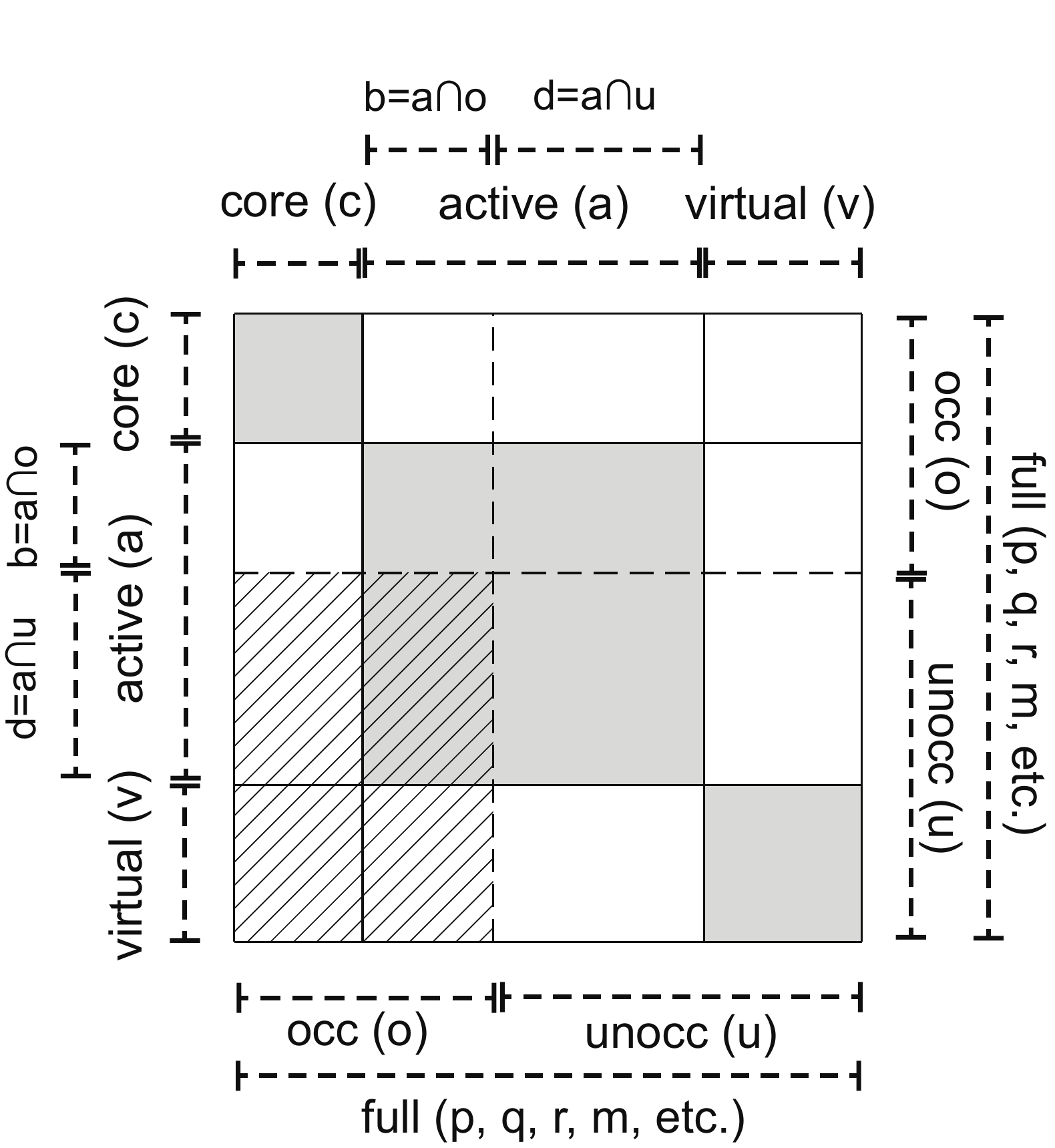}
    \caption{
    The partition of 1-electron orbitals. 
    From the view of RHF, the orbitals are classified into two parts, occupied (o) and unoccupied ones (u), divided by dashed lines. 
    From the view of CASCI, the orbitals are classified into three parts,  
    core (c, orbitals that are always fully occupied), 
    active (a, orbitals on which FCI expansion is performed), 
    and virtual (v, orbitals that are never occupied) ones, 
    divided by solid lines. 
    The gray area, i.e. $(c,c), (a,a)$ and $(v,v)$, denotes the matrix blocks where the Lagrangian matrix $I_{pq}$ is hermitian (see SI). 
    The slashed area, i.e. $(u,o)$, denotes the matrix block where independent $U_{pq}^x$ elements are solved with CPHF equations (see SI).  
    }
    \label{orb_partition}
\end{figure}
where $h_{pq}^x, (pq|rm)^x, S_{pq}^x$ are skeleton derivatives, $I_{pq}$ is the Lagrangian matrix. 
Their definitions can be found in the SI, and they can be evaluated from the atomic orbital (AO) integrals, the AO integrals' 1st derivatives, the molecular orbital (MO) coefficients and the CASCI RDMs. 
The matrix $U_{pq}^x$, however, involves first order derivatives of the MO coefficients. 
It is defined as 
\begin{equation}
    \frac{\partial C_{\mu p}}{\partial x} = \sum_q{U_{qp}^xC_{\mu q}}
    \label{U_mat}
\end{equation}
Summation condition ``$(p,q)\notin\text{gray}$'' in the last term denotes the white blocks in Fig. \ref{orb_partition}. 
It is equivalent to $(p,q)\in(a,c)\cup(v,c)\cup(v,a)\cup(c,a)\cup(c,v)\cup(a,v)$. 
The proof on Eq. \ref{force_3} can be found in SI.

$U_{pq}^x$ should be evaluated with the coupled-perturbed Hartree-Fock (CPHF) equations. 
For different orbital choices, such as RHF, ROHF, unrestricted Hartree-Fock (UHF), Kohn-Sham, or MCSCF orbitals, there are different CPHF equations, which can be derived case by case.  
However, there are some cases where $I_{pq}=I_{qp}^*$ holds for each MO pair $(p,q)$, and hence the CPHF calculation of $U_{pq}^x$ can be neglected. 
%
%
For example, this is the case in the following two occasions, where the analytic force from FCIQMC in ref. \onlinecite{Force_FCIQMC_2015} was used.

\begin{enumerate}
    \item[(1)] 
    Complete active space self-consistent field (CASSCF) is further applied on top of FCIQMC dynamics to optimize the orbitals, i.e. stochastic-MCSCF\cite{stoch_SCF_Thomas_2015}. 
    In this way, all orbitals, including the core, active and virtual orbitals, are rotated such that $I_{pq}=I_{qp}^*$ holds for each MO pair $(p,q)$. 
    \item[(2)] 
    No orbital is frozen as core or virtual, i.e. FCI calculation is performed. 
    In this way, $a$ and $v$ subspace in Fig. \ref{orb_partition} does not exist at all, and the last term in Eq. \ref{force_3} no longer appears. 
\end{enumerate}

In SI, we also present the CPHF equations for RHF and ROHF orbitals, and substitute them into the energy gradient formula. 
The Z-vector method \cite{Yamaguchi_1994} is also used to simplify the calculation. 
For RHF orbitals, the energy gradient is 
\begin{equation}
\begin{aligned}
    -F_x = \frac{\partial E_\text{tot}}{\partial x} & = \frac{\partial E_\text{nuc}}{\partial x} + \sum_{pq}{\gamma_{pq}h_{pq}^x} + \frac{1}{2}\sum_{pqrm}{\gamma_{pqrm}(pq|rm)^x} -\sum_{(p,q)\in\text{gray}}{I_{pq}S_{pq}^x} \\
    & + [(\sum_{(p,q)\in\text{slashed}}{V_{pq}B_{0,pq}^x} + \sum_{(p,q)\in(b,c)\cup(v,d)}{\frac{I_{pq}-I_{qp}^*}{\epsilon_q-\epsilon_p}B_{0,pq}^x} - \sum_{p<q,(p,q)\notin\text{gray}}{I_{pq}S_{pq}^x}) \\
    & + (\text{c.c.})]
\end{aligned}
\label{force_4_rhf}
\end{equation}
where $\epsilon_p$ is the RHF orbital energy, ``slashed'' denotes slashed region in Fig. \ref{orb_partition} (i.e. $(u,o)$), and the definitions of $V_{pq}$ and $B_{0,pq}^x$ can be found in SI. 
For ROHF orbitals, the energy gradient is the same as the one for RHF orbitals except for the definitions of $V_{pq}$, $B_{0,pq}^x$ and ``slashed'' region.
The slashed region for ROHF is shown in Fig. \ref{ROHF_orb_part}, and definitions of $V_{pq}$ and $B_{0,pq}^x$ are shown in SI Eq. 46, 47 and 49. 

\section{Computational details}

\subsection{FCIQMC simulations}

The 6-31G and cc-pVTZ basis sets were used in our calculations on N$_2$ and H$_2$O, respectively\cite{BasisSetExchange_2019}.
The molecular orbitals for the subsequent FCIQMC calculations were obtained with RHF,  performed with the PySCF package \cite{PySCF}. 
1s electrons for N and O were fully occupied and frozen, and the rest of the MOs and electrons form CAS(16o, 10e) for N$_2$ and CAS(57o, 8e) for H$_2$O. 

All FCIQMC calculations were performed with the NECI code \cite{Booth_FCIQMC_2009,Kai_NECI_2020}. 
The initiator (i-FCIQMC)\cite{Cleland_iFCIQMC_2010} and adaptive shift approach (as-FCIQMC) \cite{Ghanem_adapshift_2019, ghanem_adaptive_2020} were used in all FCIQMC calculations, with initiator threshold $n_a=3$. 
The time-step was updated using the TAU-SEARCH facility of NECI, which determines the time-step in the walker-growth stage of FCIQMC, by ensuring that the time-step is sufficiently small that no walker-blooms occur \cite{Kai_NECI_2020}.    
The semi-stochastic method \cite{petruzielo_semistochastic_2012, Blunt_semi_2015} was used, and the size of deterministic space was set to 100. 
A trial wave function was used to obtain the projected energy estimate, and the size of the trial space (composed of the most populated determinants in the ground-state wavefunction) 
was set to 10 in N$_2$ and 100 in H$_2$O. 
$D_{2h}$ point group symmetry was applied to N$_2$, and $C_1$ symmetry was used in H$_2$O. 
In calculations on H$_2$O, FCIQMC was run with different total number of walkers ($N_w$). 
To reduce initiator error, convergence of projected energy with respect to $N_w$ was reached before computing the nuclear force at each structure point. 
The convergence criterion was that the projected energy difference between the largest two successive $N_w$ is smaller than 1 mE$_\text{h}$. 
To obtain energy gradients, 1- and 2-particle reduced density matrices (1-RDM, 2-RDM) are evaluated in FCIQMC on the fly \cite{Overy_RDM_2014}. 
The energy gradients were calculated with PySCF CASCI gradient code following Eq. \ref{force_4_rhf}.

\subsection{Potential energy surface from Gaussian process regression}

Gaussian processes regression (GPR) was used to fit a smooth PES for $\text{H}_2\text{O}$ molecule. 
288 structures of H$_2$O were used in the training set. 
Both the total energy and the energy gradients w.r.t. nuclear coordinates are used as training data for the GPR model. 
%
%
Our GPR model improves upon previous GPR models designed for geometry optimisation \cite{Jonsson_GPR_2017,laude2018ab,doi:10.1063/1.5144603}.
We use $\textbf{q}=\left(\frac{1}{r_{\text{H}_1\text{H}_2}},\frac{2}{r_{\text{O}\text{H}_1}+r_{\text{O}\text{H}_2}},\frac{1}{\sqrt{r_{\text{O}\text{H}_1}r_{\text{O}\text{H}_2}}}\right)$ as the descriptor.
\textbf{q} is based on the fundamental invariants for a H$_2$O molecule \cite{derksen2002computational,doi:10.1063/1.4961454}, hence it accounts for the permutational invariance of the two H atoms. 
The training process involves solving a set of linear equations given by
\begin{equation}
(\textbf{K}_{\textbf{xx}}+\boldsymbol{\Lambda}_{\textbf{xx}})\textbf{w}_\text{\textbf{x}}=\textbf{y}_\textbf{x}.
\label{model1}
\end{equation}
Here
\begin{equation}
    \textbf{K}_\textbf{xx} = 
    \begin{pmatrix}
    \textbf{K} & \left(\frac{\text{d}\textbf{K}}{\text{d}\textbf{x}}\right)^T \\
    \frac{\text{d}\textbf{K}}{\text{d}\textbf{x}} & \frac{\text{d}}{\text{d}\textbf{x}}\left(\frac{\text{d}\textbf{K}}{\text{d}\textbf{x}}\right)^T
    \end{pmatrix}
    =
    \begin{pmatrix}
    \textbf{k}_\text{ext}^T(\textbf{x}_1)\\
    \vdots \\
    \frac{\text{d}\textbf{k}_\text{ext}^T(\textbf{x}_1)}{\text{d}\textbf{x}_1}\\
    \vdots
    \end{pmatrix}
\end{equation}
is an extension of covariance matrix 
$\textbf{K}=\left(k(\textbf{q}_i,\textbf{q}_j)\right)$
that includes derivatives of \textbf{K} with respect to the Cartesian coordinates \textbf{x}, i.e.
$
\textbf{k}_\text{ext}(\textbf{x})=
\begin{pmatrix}
k(\textbf{q}(\textbf{x});\textbf{q}_1);...;
& \frac{\text{d}k(\textbf{q}(\textbf{x}),\textbf{q}_1)}{\text{d}\textbf{x}_1}~;...
\end{pmatrix}.
$
The Gaussian kernel 
$k(\textbf{q}_i,\textbf{q}_j)=\mathrm{exp}\left(-\frac{(\textbf{q}_i-\textbf{q}_j)^2}{\sigma^2}\right)$
is used.
$\boldsymbol{\Lambda}_\textbf{xx}$ is the noise matrix in Cartesian coordinates 
$\begin{pmatrix}
\sigma_{V}^2\textbf{I} &  \\
 & \sigma_{g}^2\textbf{I}
\end{pmatrix}$.
$\textbf{y}_\textbf{x}=\left(V(\textbf{x}_1)-\bar{V},...,\left.\frac{\text{d}V}{\text{d}\textbf{x}}\right|_{\textbf{x}_1},...\right)^T$ is the training data and $\bar{V}$ is the average potential energy of the training data. 


The prediction of the energy and Cartesian force of a new geometry $\textbf{x}^*$ is given by
\begin{equation}
\begin{pmatrix}
V(\textbf{x}^*)-\bar{V}\\\left.\frac{\text{d}V}{\text{d}\textbf{x}}\right|_{\textbf{x}^*}
\end{pmatrix}
=
\begin{pmatrix}
\textbf{k}_\text{ext}^T(\textbf{x}^*)\\
\frac{\text{d}\textbf{k}_\text{ext}^T(\textbf{x}^*)}{\text{d}\textbf{x}^*}
\end{pmatrix}
\textbf{w}_\textbf{x}.
\end{equation}

\section{Results and Discussions}

In this section, we present our results on the potential energy curve of N$_2$, and PES of a water molecule. 

\subsection{N$_2$}

\begin{figure}
    \centering
    \includegraphics[width=16.0cm]{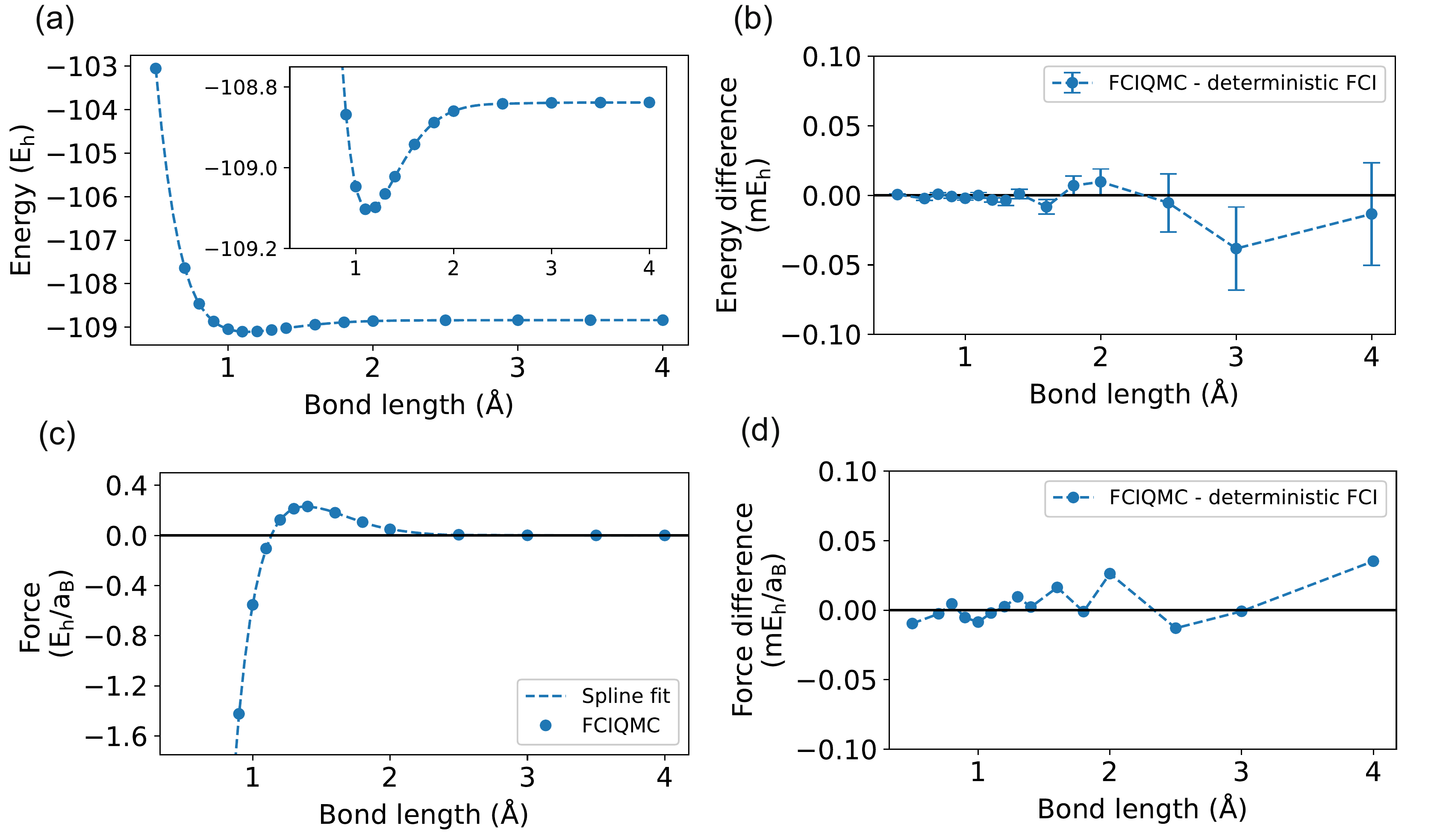}
    \caption{
    Binding curves of N$_2$. 
    (a) FCIQMC projected energy, as a function of bond length. The dashed line is a cubic spline fit. The inset shows a zoom-in view near the equilibrium.  The errorbars of FCIQMC energy are too small to be seen on this scale. 
    (b) Total energy difference between the FCIQMC projected estimate and exact FCI. 
    (c) FCIQMC analytic energy gradient. The dots are from FCIQMC. The dashed line is from the derivative of the cubic spline fit of the total energy. 
    (d) Energy gradient difference between FCIQMC and exact FCI. Both values are from analytic gradient formula. 
    }
    \label{N2}
\end{figure}

In N$_2$ calculation, we consider a (16o,10e) CAS to perform FCIQMC and exact FCI calculation within CAS. 
The binding curve of N$_2$ is plotted in Fig. \ref{N2}. 
FCIQMC total energies are plotted in (a), and the continuous binding curve (plotted with dashed line) is fitted with cubic spline. 
Analytic energy gradient from FCIQMC is plotted in (c), and is compared with the numerical force, i.e. the derivative of the cubic spline energy curve (plotted with dashed line). 
We see that the analytic force is very close to the numerical force, indicating that our analytic gradient formula is accurate and applies to FCIQMC.  
The difference between FCIQMC and exact FCI is compared in (b) and (d). 
The error of FCIQMC total energy is within 0.05 mE$_\text{h}$ to the deterministic FCI value, and the force error is within 0.05 $\text{mE}_\text{h}/\text{a}_\text{B}$. 
Both of these errors are negligible in nuclear motion calculations, such as structure optimization and molecular dynamics. 
The small difference between FCIQMC and exact FCI indicates small initiator error and high accuracy of FCIQMC. 

As discussed in Section \ref{theory}, the addition term introduced by the frozen core approximation is the final term in Eq. \ref{force_3}. 
Therefore, we can split the total force into the following two terms, and compare their contributions to see the importance of the additional force introduced by the frozen core. 
\begin{equation}
\begin{aligned}
    \text{Term 1} & = \frac{\partial E_\text{nuc}}{\partial x} + \sum_{pq}{\gamma_{pq}h_{pq}^x} + \frac{1}{2}\sum_{pqrm}{\gamma_{pqrm}(pq|rm)^x} - \frac{1}{2}\sum_{pq}{S_{pq}^x(I_{pq}+I_{qp}^*)} \\
    \text{Term 2} & = \frac{1}{2}\sum_{(p,q)\notin(c,c)\cup(b,b)\cup(d,d)\cup(v,v)} {(I_{pq}-I_{qp}^*)(U_{pq}^x-U_{qp}^{x*})}
\end{aligned}
\label{Term1_2}
\end{equation}
Term 1 is identical to nuclear force formula in all-electron FCI case. 
Term 2 is equivalent to the final term in Eq. \ref{force_3} on the assumption of $I_{pq}=I_{qp}^*$, $(p,q)\in(b,d)$, which can be proved to be true with CASCI variational condition (See SI). 

\begin{figure}
    \centering
    \includegraphics[width=16.0cm]{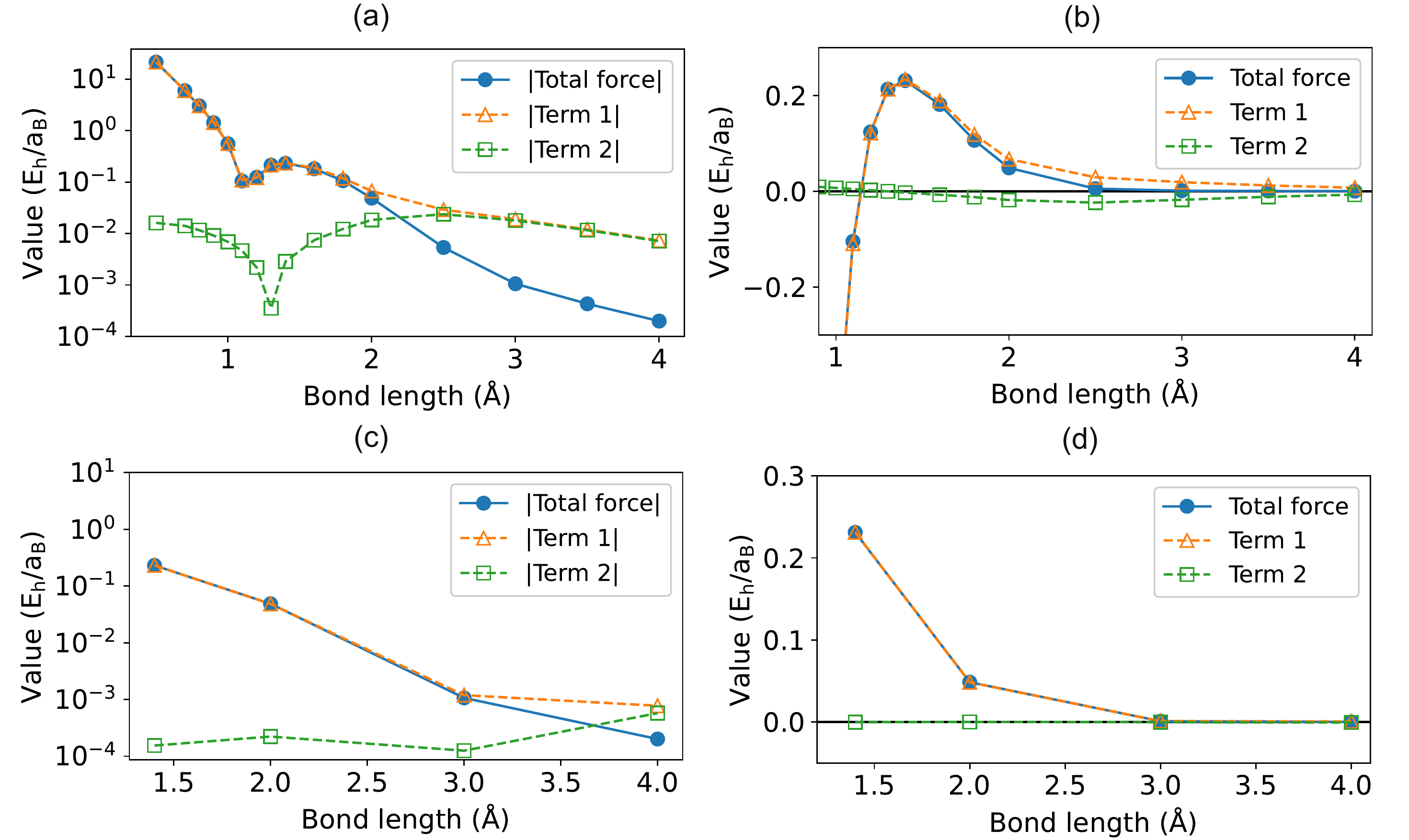}
    \caption{Term 1, Term 2 and total nuclear force for (a,b) the frozen core case and (c,d) the all-electron case. 
    In (a,c), the absolute values are plotted with logarithmic coordinate, while in (b,d) the original values are plotted with linear coordinate.  
    Term 1 and Term 2 are defined in Eq. \ref{Term1_2}. }
    \label{N2_2term_comp}
\end{figure}

Fig. \ref{N2_2term_comp} shows the decomposition of the force as a function of bond length of N$_2$. 
In the frozen core case (Fig. \ref{N2_2term_comp}a), the proportion between Term 1 and Term 2 behaves differently at small and large bond length. 
At small bond lengths, Term 2 is 2-3 orders of magnitude smaller than Term 1, hence it has negligible effect on the total nuclear force. 
At large bond lengths, Term 2 becomes comparable to Term 1 with the same order of magnitude. 
Such a transition occurs around $r=2.0$ \AA, where the molecule starts to dissociate. 
This feature can be understood as follows. 
At large bond lengths, the ground state is poorly described with RHF orbitals, which is more ``different'' from CASSCF orbitals than at small bond lengths. 
The large difference between RHF and CASSCF orbitals leads to large Term 2 at large bond lengths. 

In the all-electron case (Fig. \ref{N2_2term_comp}b), Term 2 should be strictly zero assuming that $I_{pq}=I_{qp}^*$ for $(p,q)\in(b,d)$. 
However, in FCIQMC, the initiator error and stochastic fluctuation of wave function from the ground state introduces some deviation from this identity, and therefore lead to a non-zero Term 2. 
In our numerical tests, molecular geometries with different stretch is chosen, and FCIQMC calculation is performed with a maximum of 2 million walkers. 
These numerical tests show that CASCI variational condition is achieved approximately such that Term 2 makes small contribution to the total force on the order of $10^{-4}$ $\text{E}_\text{h}/\text{a}_\text{B}$. 
This contribution is sufficiently small for further exploration on PES. 
However, our tests suggest that computing Term 2 in the total force would be an additional validation of all-electron FCIQMC convergence. 

\subsection{H$_2$O}

The water monomer PES is often used for studies of water splitting\cite{Dixon_water_science_1999,vanHarrevelt_pdwater_2000,Chang_water_nature_2021} and infrared spectrum\cite{Polyansky_2003,Barletta_2006,Bubukina_2011,Mizus_IR_2018,Exomol_water_2018}, etc. 
The triatomic H$_2$O molecule has in total 3 internal degrees of freedom, which can be defined with parameters $(r_1, r_2, \theta)$. 
$r_1$ and $r_2$ denote the two O-H bond lengths, while $\theta$ denotes the bond angle H-O-H. 
In this section, we use the total energy and energy gradients of 288 structures of an H$_2$O monomer as the training set, and obtain a smooth PES with GPR model. 
%

\begin{figure}
    \centering
    \includegraphics[width=16.0cm]{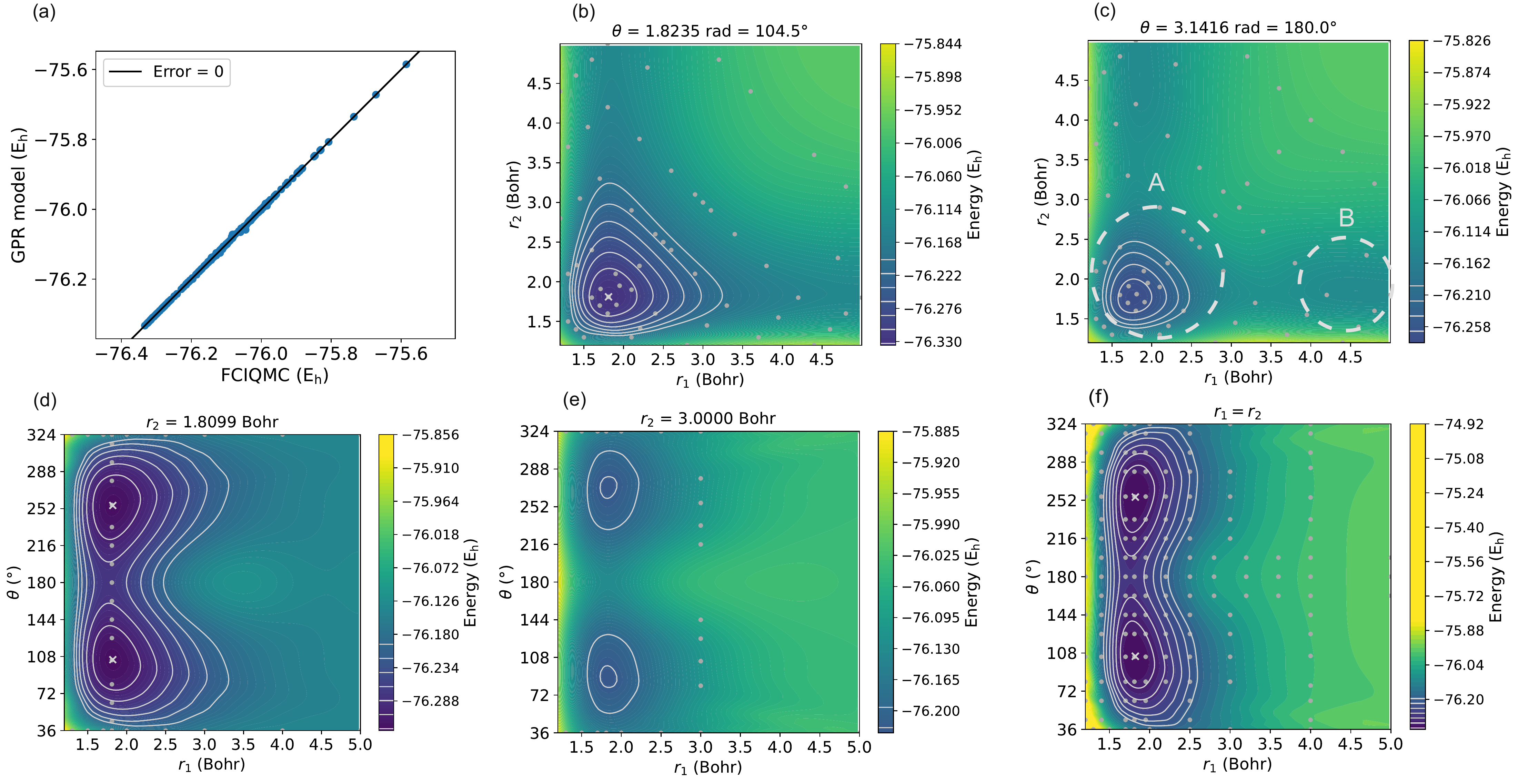}
    \caption{
    (a) GPR model energy versus FCIQMC energy.
    (b-f) Contour plot of PES on the following 2-dimensional sections. 
    (b) $\theta=1.8235=104.5\degree$;
    (c) $\theta=\pi=180\degree$;
    (d) $r_2=1.8099\text{ Bohr}$; 
    (e) $r_2=3.0000\text{ Bohr}$; 
    (f) $r_1=r_2$. 
    The PES sections are shown within the following conditions: $1.2\text{ Bohr }\leq r_1, r_2\leq 5\text{ Bohr}$, $36\degree\leq\theta\leq180\degree$.
    The experimental equilibrium geometry is marked with a cross, and FCIQMC points are marked with round dots. 
    The solid lines represent isosurfaces 5000, 10000, 15000, 20000, 25000 or 30000 cm$^{-1}$ above the global minimum.
    }
    \label{h2o}
\end{figure}

Fig. \ref{h2o}a plots the GPR energy versus the computed FCIQMC energy, where the root mean square (RMS) error is 1.5 mE$_\text{h}$, 1.3\% of the standard deviation of energy data points (119 mE$_\text{h}$). 
Overall, our GPR model reproduces the PES with high quality, and the erroneous points mainly come around the so-called conical crossing point, between the ground state $\tilde{X} ^1A'$ and excited state $\tilde{B} ^1A'$\cite{Dixon_water_science_1999, Chang_water_nature_2021}. 
Conical crossing leads to discontinuity in energy gradient and increases the fitting error in GPR.  
Apart from the conical crossing regime, the negligible error suggests that our GPR model can reproduce almost exactly the PES from FCIQMC energy and force data.
In addition, if FCIQMC energies and forces would have had large statistical errors, the GPR model would have large fitting errors. 
Therefore, the high quality of our GPR PES also suggests that the FCIQMC energy and force calculations are well converged, and the resultant ground state PES can be used for further applications.
%
%
More discussions on conical crossing is presented in the SI, and we will also see its main feature on the ground state PES presented below.

Different 2-dimensional projections of the PES are shown as contours in Fig. \ref{h2o}(b-f).
Fig. \ref{h2o}b is an iso-$\theta$ cut of PES at $\theta=104.5\degree$, near the equilibrium bond angle of a water monomer.
In the equilibrium bond angle section, the PES has only one global minimum at (1.821 Bohr, 1.821 Bohr, 105.0$\degree$). 
The GPR global minimum lies close to the experimental global minimum (1.810 Bohr, 1.810 Bohr, 104.5$\degree$\cite{cccbdb}), which is marked in Fig. \ref{h2o} (b) with cross. 
Fig. \ref{h2o}c is another iso-$\theta$ cut at $\theta=180\degree$, i.e. with the linear H-O-H water molecule.
At the linear geometry (Fig. \ref{h2o}c), however, there is a barrier ($r_1\sim3.5$ Bohr, $r_2\sim2.0$ Bohr) lying between minimum A and minimum B. 
The barrier stems from the conical crossing between $\tilde{X} ^1A'$ and $\tilde{B} ^1A'$, alluded to above, which is hard to be described accurately with a smooth GPR fit. 

In Fig. \ref{h2o} (d) and (e), one of the two O-H bonds are fixed at 1.8099 Bohr (equilibrium O-H bond length in water) in (d), and at 3.0 Bohr (a stretched O-H bond length) in (e), and the other hydrogen atom is allowed to move without restriction. 
On the $r_2=1.8099$ Bohr section, there are two valleys on the PES symmetric with respect to the $\theta=180\degree$ line. 
The two valleys represent the same global minimum, and can be transformed into each other by a rotation. 
The local maximum near the linear geometry ($r_1\sim3.5$ Bohr, $\theta\sim180\degree$) also results from the conical crossing point. 
In Fig. \ref{h2o}f, the two O-H bonds are kept at the same length, i.e. $r_1=r_2$, and the double-well feature similar to Fig. \ref{h2o}d occurs. 
In all of these sections (Fig. \ref{h2o} (b-f)), we also show six iso-energy lines with excitation energy $<30,000$ cm$^{-1}$.
In practical applications, the PES of the water monomer is often used to calculate rotation-vibrational levels with excitation energy $\lessapprox10,000$ cm$^{-1}$, which lies within a small area near the global minimum of our PES and is quite far away from the conical crossing point.
The near-FCI nature of this PES, as well as the correct shape in the rotation-vibration involved region, implies the possibility to study rotation-vibration levels with little CI truncation error. 
However, the quantitative study of rotation-vibration levels is beyond the scope of this work, and is left for future research.

\section{Conclusions}

To conclude, this work discusses how to obtain nuclear forces from FCIQMC and more generally stochastic CASCI calculations. 
Our method supplements previous work, and can be extended to different scenarios, e.g. employing frozen cores, and using a wide range of orbitals.
We also clarify the different contributions of the total FCIQMC force, especially the part that has been neglected in previous work.
As an illustration of our method, we produce a full dimensional water molecule PES at the FCIQMC level, which can be further used for computing rotation-vibration spectrum of water and studying dynamic effects in water splitting.
Overall, the method will enable us to obtain high quality PESs from FCIQMC for molecules and materials.

\section{Supplementary information}

\makeatletter 
\renewcommand{\thefigure}{S\@arabic\c@figure}
\renewcommand{\theequation}{S\arabic{equation}}
\setcounter{figure}{0}
\setcounter{equation}{0}
\makeatother

\subsection{Analytic force in FCIQMC}

\subsubsection{FCIQMC method}

In FCIQMC, one seeks a stochastic FCI solution to the \textit{ab initio} Hamiltonian.  \cite{Booth_FCIQMC_2009,Kai_NECI_2020}. 
\begin{equation}
    \hat{H} = E_\text{nuc} + \sum_{pq}{h_{pq}\hat{p}^\dagger \hat{q}} + \frac{1}{2}\sum_{pqrm}{(pq|rm)\hat{p}^\dagger \hat{r}^\dagger \hat{m}\hat{q}}
    \label{ab_initio_Hamil}
\end{equation}
where $E_\text{nuc}$ denotes the nuclear repulsion term, $p,q,r,m$ denote molecular orbital (MO) indices, and $h_{pq}$ and $(pq|rm)$ denote the 1-body and 2-body integrals, respectively.
The 2-body integrals are written in chemists' notations. Namely, 
\begin{equation}
    (pq|rm)=\int{p^*(x_1)r^*(x_2)\frac{1}{|x_1-x_2|}q(x_1)m(x_2)\text{d}x_1\text{d}x_2}
\end{equation}

As in deterministic FCI, FCIQMC solves the eigenvalue problem of $\hat{H}$ in N-electron Hilbert space under the Slater determinant (SD) basis of MO. 
The ground state FCI wave function, formulated as
\begin{equation}
    |\Psi_\text{gs}\rangle = \sum_I{C_I|D_I\rangle}, 
\end{equation}
is determined by the eigenvalue equation as follows. 
\begin{equation}
    \sum_J{H_{IJ}C_J} = E_\text{tot}C_I
    \label{variational}
\end{equation}
$I,J$ denote SD indices, $C_I$ and $H_{IJ}$ denote the CI coefficient of $|D_I\rangle$ and the Hamiltonian matrix element between $|D_I\rangle$ and $|D_J\rangle$. 
\begin{equation}
\begin{aligned}
    C_I = \langle D_I|\Psi_\text{gs}\rangle, \qquad
    H_{IJ} = \langle D_I|\hat{H}|D_J\rangle
\end{aligned}
\end{equation}

In FCIQMC the eigenvalue problem is solved by a stochastic projection algorithm. 
The corresponding projection operator $\hat{P}$ is defined as Eq. \ref{projector_p}. 
One can prove that performing $\hat{P}$ repeatedly on an arbitrary wavefunction with non-zero overlap with ground state yields ground state wavefunction (Eq. \ref{projector_7}). 
\begin{equation}
    \hat{P} = \hat{\mathbbm{1}} - \delta\tau(\hat{H}-S\hat{\mathbbm{1}})
    \label{projector_p}
\end{equation}
\begin{equation}
    \lim_{n\rightarrow\infty}{\hat{P}^n|\Psi_0\rangle} \propto |\Psi_\text{gs}\rangle, \qquad \text{if} \quad \langle\Psi_0|\Psi_\text{gs}\rangle \neq 0
    \label{projector_7}
\end{equation}
Therefore, one can start with an arbitrary wave function, e.g. Hartree-Fock (HF) wavefunction, and perform $\hat{P}$ on it repeatedly until the wavefunction converges. 
However, as in Lanczos diagonalization, this projection method is also hindered by the combinatorial rise of Hilbert space size (i.e. the number of all possible SDs). 

In FCIQMC, the CI coefficients are represented with a population of walkers, and the linear operator $\hat{P}$ are stochastically simulated by three successive steps, spawning, cloning or death, and annihilation, in each iteration. 
In most \textit{ab initio} systems, only a small number of SDs dominate the ground state wavefunction, and the majority of SDs have very small amplitude. 
%
%
In this way, the distribution of walkers can reach equilibrium rather quickly to represent the population in the Hilbert space, which is also a representation of the ground state wavefunction. 

\subsubsection{General formula of FCIQMC force}

The energy gradients for HF, CI and MCSCF have been discussed in Ref \onlinecite{Yamaguchi_1994}.
However, there are some limitations and the formulations are only applicable with certain conditions.
Here we discuss the compete formulations of general analytic force of FCIQMC and the simplifications that can be used in special scenarios. 
%
%

%
In general, one can consider a complete active space problem where several orbitals and electrons are selected within the entire 1-electron orbital space.
%
In this way, the 1-electron orbital space can be divided into three parts: core ($c$), active ($a$) and virtual ($v$) space. 
FCI diagonalization can be performed within the active space to generate a CASCI wavefunction. 

The total energy of CASCI wavefunction is formulated as follows. 
\begin{equation}
    E_\text{tot} = E_\text{nuc} + \sum_{pq}{\gamma_{pq}h_{pq}} + \frac{1}{2}\sum_{pqrm}{\gamma_{pqrm}(pq|rm)}
    \label{total_ene}
\end{equation}
where $\gamma_{pq}$ and $\gamma_{pqrs}$ represent the one and two-particle reduced density matrices (1-RDM, 2-RDM), respectively. 
\begin{equation}
\begin{aligned}
    \gamma_{pq} & = \langle\Psi_\text{gs}|\hat{p}^\dagger \hat{q}|\Psi_\text{gs}\rangle 
    = \sum_{IJ}{C_I^* C_J\langle D_I|\hat{p}^\dagger \hat{q}|D_J\rangle} \\
    \gamma_{pqrm} & = \langle\Psi_\text{gs}|\hat{p}^\dagger \hat{r}^\dagger \hat{m}\hat{q}|\Psi_\text{gs}\rangle 
    = \sum_{IJ}{C_I^* C_J\langle D_I|\hat{p}^\dagger \hat{r}^\dagger \hat{m}\hat{q}|D_J\rangle} \\
\end{aligned}
\label{RDM}
\end{equation}
The force, defined as the negative gradient of the total energy $E_\text{tot}$, are evaluated by calculating the first order derivative of Eq. \ref{total_ene}. 
\begin{equation}
\begin{aligned}
    -F_x = \frac{\partial E_\text{tot}}{\partial x} = \frac{\partial E_\text{nuc}}{\partial x} + \sum_{pq}{\gamma_{pq}\frac{\partial h_{pq}}{\partial x}} + \frac{1}{2}\sum_{pqrm}{\gamma_{pqrm}\frac{\partial (pq|rm)}{\partial x}}
\end{aligned}
\label{force_1}
\end{equation}
where $x$ is the nuclear degree of freedom of the system. 
The 2nd equality follows the identity of RDM derivatives due to the variational condition.
\begin{equation}
    \sum_{pq}{h_{pq}\frac{\partial \gamma_{pq}}{\partial x}} + \frac{1}{2}\sum_{pqrm}{(pq|rm)\frac{\partial \gamma_{pqrm}}{\partial x}} = 0
    \label{RDM_deriv_id}
\end{equation}
This is because
\begin{equation}
\begin{aligned}
    & \sum_{pq}{h_{pq}\frac{\partial \gamma_{pq}}{\partial x}} + \frac{1}{2}\sum_{pqrm}{(pq|rm)\frac{\partial \gamma_{pqrm}}{\partial x}} \\
    = & \sum_{IJ}{\frac{\partial (C_I^*C_J)}{\partial x} (\sum_{pq}{h_{pq}\langle D_I|\hat{p}^\dagger \hat{q}|D_J\rangle} + \sum_{pqrm}{(pq|rm)\langle D_I|\hat{p}^\dagger \hat{r}^\dagger \hat{m}\hat{q}|D_J\rangle)}} \\
    = & \sum_{IJ}{(\frac{\partial C_I^*}{\partial x}C_J + C_I^*\frac{\partial C_J}{\partial x}) \langle D_I|(\hat{H} - E_\text{nuc}\hat{\mathbbm{1}})|D_J\rangle} \\
    = & (E_\text{tot}-E_\text{nuc}) \sum_I{\frac{\partial C_I^*C_I}{\partial x}} \\
    = & (E_\text{tot}-E_\text{nuc}) \frac{\partial\langle\Psi_\text{gs}|\Psi_\text{gs}\rangle}{\partial x} = 0
    \label{RDM_deriv_id_proof}
\end{aligned}
\end{equation}
The 3rd equality of Eq. \ref{RDM_deriv_id_proof} follows the variational condition (Eq. \ref{variational}), and the 5th equality follows the normalization condition $\langle\Psi_\text{gs}|\Psi_\text{gs}\rangle =1$.

The first order derivatives of 1-electron integrals, namely $\frac{\partial h_{pq}}{\partial x}$, can be calculated as follows. 
\begin{equation}
\begin{aligned}
    \frac{\partial h_{pq}}{\partial x} & = \sum_{\mu\nu}{(\frac{\partial h_{\mu\nu}}{\partial x}C_{\mu p}^*C_{\nu q} + h_{\mu\nu}\frac{\partial C_{\mu p}^*}{\partial x} C_{\nu q} + h_{\mu\nu}C_{\mu p}^*\frac{\partial C_{\nu q}}{\partial x})} \\
    & = h_{pq}^x + \sum_r{(h_{pr}U_{rq}^x + U_{rp}^{x*}h_{rq})}
\end{aligned}
\label{h_MO_deriv}
\end{equation}
Here $h_{pq}^x$ are skeleton derivatives. 
$U_{pq}^x$ are related to MO coefficient derivatives. 
\begin{equation}
\begin{aligned}
    h_{pq}^x & = \sum_{\mu\nu}{\frac{\partial h_{\mu\nu}}{\partial x}C_{\mu p}^*C_{\nu q}} \\ 
    (pq|rm)^x & = \sum_{\mu\nu\rho\sigma}{\frac{\partial (\mu\nu|\rho\sigma)}{\partial x}C_{\mu p}^*C_{\nu q}C_{\rho r}^*C_{\sigma m}}
\end{aligned}
\end{equation}
\begin{equation}
    \frac{\partial C_{\mu p}}{\partial x} = \sum_q{U_{qp}^xC_{\mu q}}
    \label{U_mat}
\end{equation}
Similarly, 
\begin{equation}
    \frac{\partial (pq|rm)}{\partial x} = (pq|rm)^x + \sum_t{((pt|rm)U_{tq}^x + (pq|rt)U_{tm}^x + (tq|rm)U_{tp}^{x*} + (pq|tm)U_{tr}^{x*})}
\label{g_MO_deriv}
\end{equation}

Substituting Eq. \ref{h_MO_deriv} and Eq. \ref{g_MO_deriv} into Eq. \ref{force_1}, we have
\begin{equation}
     -F_x = \frac{\partial E_\text{tot}}{\partial x} = \frac{\partial E_\text{nuc}}{\partial x} + \sum_{pq}{\gamma_{pq}h_{pq}^x} + \frac{1}{2}\sum_{pqrm}{\gamma_{pqrm}(pq|rm)^x} + \sum_{pq}{(I_{pq}U_{pq}^x+I_{qp}^*U_{qp}^{x*})}
     \label{force_2}
\end{equation}
where $I_{pq}$ is the Lagrangian matrix for FCI wave function. 
\begin{equation}
    I_{pq} = \sum_r{h_{rp}\gamma_{rq}} + \sum_{trm}{(tp|rm)\gamma_{tqrm}}
\end{equation}
From the orthonormal condition for MO, one can prove that
\begin{equation}
\begin{aligned}
    U_{pq}^x + U_{qp}^{x*} + S_{pq}^x = 0 \\
    S_{pq}^x = \sum_{\mu\nu}{\frac{\partial S_{\mu\nu}}{\partial x}C_{\mu p}^*C_{\nu q}}
\end{aligned}
\label{Upq_Spqx}
\end{equation}
Substituting Eq. \ref{Upq_Spqx} into Eq. \ref{force_2}, one has
\begin{equation}
\begin{aligned}
    -F_x = \frac{\partial E_\text{tot}}{\partial x} & = \frac{\partial E_\text{nuc}}{\partial x} + \sum_{pq}{\gamma_{pq}h_{pq}^x} + \frac{1}{2}\sum_{pqrm}{\gamma_{pqrm}(pq|rm)^x} \\
    & - \frac{1}{2}\sum_{pq}{S_{pq}^x(I_{pq}+I_{qp}^*)} + \frac{1}{2}\sum_{pq}{(I_{pq}-I_{qp}^*)(U_{pq}^x-U_{qp}^{x*})}
    \label{force_3}
\end{aligned}
\end{equation}

All quantities in Eq. \ref{force_3} can be directly evaluated from MO coefficients and atomic orbital (AO) integrals, 
except $U_{pq}^x$, and therefore the final term in Eq. \ref{force_3}.
$U_{pq}^x$ can be evaluated with coupled-perturbed Hartree-Fock (CPHF) equations. 
However, there are some cases where $I_{pq}=I_{qp}^*$ holds for each MO pair $(p,q)$, and hence calculation of $U_{pq}^x$ is not needed.
This include the following two scenarios.

\begin{enumerate}
    \item[(1)] 
    Complete active space self-consistent field (CASSCF) is further applied on top of FCIQMC dynamics to optimize the orbitals, i.e. stochastic-MCSCF\cite{stoch_SCF_Thomas_2015,Kai_NECI_2020}. 
    %
    In this way, all orbitals, including the core, active and virtual orbitals, are rotated such that $I_{pq}=I_{qp}^*$ holds for each MO pair $(p,q)$. 
    \item[(2)] 
    No orbital is frozen as core or virtual. 
    Namely, FCI calculation is performed. 
\end{enumerate}

\begin{figure}
    \centering
    \includegraphics[width=8.0cm]{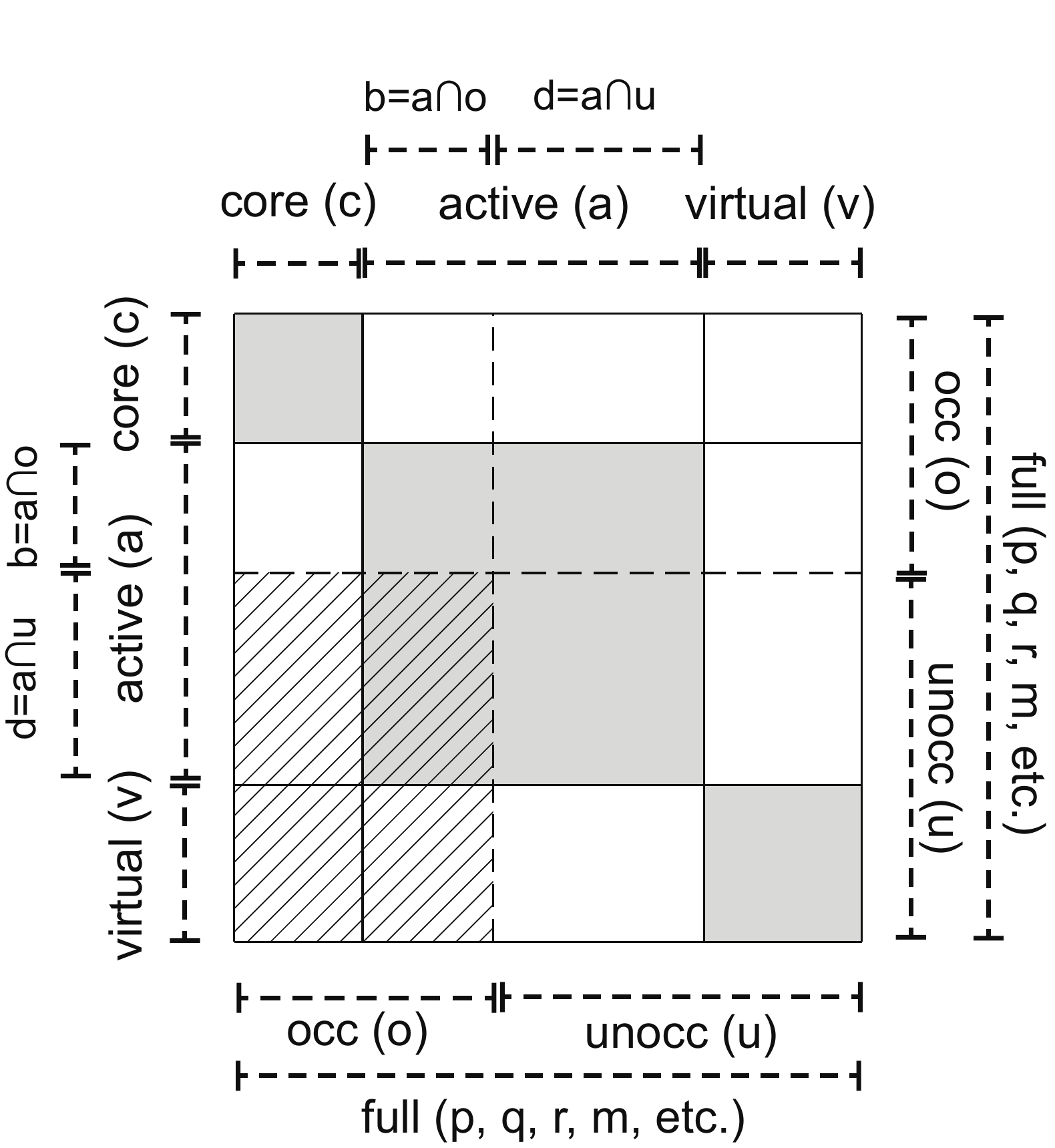}
    \caption{
    The partition of 1-electron RHF orbitals to construct the Lagrangian matrix $I_{pq}$. 
    From the view of RHF, the orbitals are classified into two parts, occupied (o) and unoccupied ones (u), divided by dashed lines. 
    From the view of CASCI, the orbitals are classified into three parts,  
    core (c, orbitals that are always fully occupied), 
    active (a, orbitals on which FCI expansion is performed), 
    and virtual (v, orbitals that are never occupied) ones, 
    divided by solid lines. 
    The gray area, i.e. $(c,c), (a,a)$ and $(v,v)$, denotes the matrix blocks where the Lagrangian matrix $I_{pq}$ is hermitian, and in which the orbital rotations do not change the wavefunction.  
    The slashed area, i.e. $(u,o)$, denotes the matrix block where independent $U_{pq}^x$ elements are solved with CPHF equations.  
    }
    \label{orb_partition_SI}
\end{figure}

To prove the hermiticity of $I_{pq}$ in case (2) (FCI case), we look at the partition of orbitals in Fig. \ref{orb_partition_SI}. 
Actually, one can prove that in general CASCI, the (core,core), (active,active), (virtual,virtual) (gray area in Fig. \ref{orb_partition_SI}. $(c,c)$, $(a,a)$, $(v,v)$ for short) blocks of $I_{pq}-I_{qp}^*$ are zero.

\textit{Proof} Suppose that $R$ is a block-diagonal unitary transformation with non-zero elements only in $(c,c), (a,a)$ and $(v,v)$ region. 
The rotation of MO coefficients are formulated as follows. 
\begin{equation}
    R: C_{\mu p} \stackrel{R}{\longrightarrow} C_{\mu p}' = \sum_q{R_{pq}C_{\mu q}}
\end{equation}
The FCI wave function is invariant under basis rotation within the CAS space, i.e. $|\Psi_\text{gs}\rangle$ is invariant under $R$. 
Therefore, the 1-RDM is transformed as follows under $R$.
\begin{equation}
    R: \gamma_{pq} \stackrel{R}{\longrightarrow} \gamma_{pq}' = \sum_{p'q'}{R_{pp'}R_{qq'}^*\gamma_{p'q'}}
\end{equation}
Then the transformation of derivatives of RDM, $\frac{\partial \gamma_{pq}}{\partial x}$, under $R$, noting that $R$ is relevant to nuclear coordination $x$, is shown in Eq. \ref{dm1_rot_global}. 
\begin{equation}
\begin{aligned}
    \frac{\partial \gamma_{pq}}{\partial x} \stackrel{R}{\longrightarrow} \frac{\partial \gamma_{pq}'}{\partial x} 
    = \sum_{p'q'}{(R_{pp'}R_{qq'}^* \frac{\partial \gamma_{p'q'}}{\partial x} 
    + \frac{\partial R_{pp'}}{\partial x}R_{qq'}^*\gamma_{p'q'}
    + R_{pp'}\frac{\partial R_{qq'}^*}{\partial x}\gamma_{p'q'})}
\end{aligned}
\label{dm1_rot_global}
\end{equation}
If $R$ is identity at $x=x_0$ (i.e. $R(x_0)=I$), and $T_{pq}=\frac{\partial R_{pq}}{\partial x}\bigg|_{x=x_0}$, then 
\begin{equation}
\begin{aligned}
    \frac{\partial \gamma_{pq}}{\partial x}\bigg|_{x=x_0} \stackrel{R}{\longrightarrow} \frac{\partial \gamma_{pq}'}{\partial x}\bigg|_{x=x_0} 
    & = \frac{\partial \gamma_{pq}}{\partial x}\bigg|_{x=x_0} + \sum_r{(T_{pr}\gamma_{rq}-\gamma_{pr}T_{rq})}
\end{aligned}
\end{equation}
Similarly, 
\begin{equation}
\begin{aligned}
    \frac{\partial \gamma_{pqrm}}{\partial x}\bigg|_{x=x_0} \stackrel{R}{\longrightarrow} \frac{\partial \gamma_{pqrm}'}{\partial x}\bigg|_{x=x_0} 
    & = \frac{\partial \gamma_{pqrm}}{\partial x}\bigg|_{x=x_0} + \sum_t{(T_{pt}\gamma_{tqrm}+T_{rt}\gamma_{pqtm}-\gamma_{ptrm}T_{tq}-\gamma_{pqrt}T_{tm})}
\end{aligned}
\label{RDM_deriv_trans}
\end{equation}
Now return to the identity Eq. \ref{RDM_deriv_id}.
It comes from the variational condition of FCI and CASCI, and should hold upon transformation under $R$. 
Substitute Eq. \ref{RDM_deriv_trans} into the transformation of Eq. \ref{RDM_deriv_id} under $R$, and one gets
\begin{equation}
\begin{aligned}
    & \sum_{pq}{h_{pq}\frac{\partial \gamma_{pq}}{\partial x}} + \frac{1}{2}\sum_{pqrm}{(pq|rm)\frac{\partial \gamma_{pqrm}}{\partial x}} \\
    \stackrel{R}{\longrightarrow} & \sum_{pq}{h_{pq}\frac{\partial \gamma_{pq}}{\partial x}} + \frac{1}{2}\sum_{pqrm}{(pq|rm)\frac{\partial \gamma_{pqrm}}{\partial x}} + \sum_{pq}{T_{pq}(I_{pq}^*-I_{qp})} \\
    & \Rightarrow \sum_{pq}{T_{pq}(I_{pq}^*-I_{qp})} = 0
\end{aligned}
\end{equation}
Since $T_{pq}$ can be any anti-hermitian block-diagonal matrix in $(c,c),(a,a)$ and $(v,v)$, 
one has $I_{pq}=I_{qp}^* \quad \forall (p,q)\in(c,c),(a,a)\text{ or }(v,v)$. $\qedsymbol$

Therefore, Eq. \ref{force_3} can be simplified as 
\begin{equation}
\begin{aligned}
    -F_x = \frac{\partial E_\text{tot}}{\partial x} & = \frac{\partial E_\text{nuc}}{\partial x} + \sum_{pq}{\gamma_{pq}h_{pq}^x} + \frac{1}{2}\sum_{pqrm}{\gamma_{pqrm}(pq|rm)^x} \\
    & - \frac{1}{2}\sum_{pq}{S_{pq}^x(I_{pq}+I_{qp}^*)} + \frac{1}{2}\sum_{(p,q)\notin \text{gray}}{(I_{pq}-I_{qp}^*)(U_{pq}^x-U_{qp}^{x*})}
\end{aligned}
\label{force_4}
\end{equation}
where $(p,q)\notin\text{gray}$ means that $(p,q)$ lies in the white blocks in Fig. \ref{orb_partition_SI} (i.e. $(a,c), (v,c), (v,a), (c,a), (c,v)$ and $(a,v)$). 
In FCI without frozen core, the $(a,a)$ covers the entire 1-electron space.
Thus we don't need to calculate $U^x$ in FCI case. 

However, if there are frozen orbitals, then it is necessary to determine the off-diagonal blocks of $U^x$. 
For different orbital schemes, we must calculate the MO coefficient derivative case by case. 
This can be explained as follows. 
If there are frozen core orbitals, then the wavefunction and the ground state energy must depend on the frozen core choice. 
We can assume at nuclear coordinate $x$ the energy is $E[A(x);C(x)]$, where $A(x)$ and $C(x)$ denote the atomic integrals and the MO coefficients at $x$, respectively. 
Then for different $\frac{\text{d}C}{\text{d}x}$, namely the MO coefficient derivatives, we have different $C(x+\delta x)$, and thereby different $E[A(x+\delta x),C(x+\delta x)]$ and different energy gradient w.r.t. $x$.
Therefore, nuclear force for FCIQMC with frozen core and virtual orbitals must depend on $\frac{\text{d}C}{\text{d}x}$, which appears only in the final term in Eq. \ref{force_4}. 

In the following two sub-sections we revisit energy gradients for stochastic CASCI with RHF orbitals, and present the energy gradient formula for stochastic CASCI with ROHF orbitals. 

\subsubsection{Formula for RHF orbitals}

RHF orbitals are determined with RHF SCF equations. 
\begin{equation}
    F_{pq} = h_{pq} + \sum_o{[2(pq|oo)-(po|oq)]} = \epsilon_p \delta_{pq}
    \label{RHF_SCF}
\end{equation}
where $p,q$ denote spatial orbitals, and $o$ denotes the occupied orbitals (see Fig. \ref{orb_partition_SI}). 
In this and the next subsection, the orbital indices $c,a,v,b,d,o,u$ denote orbitals in the corresponding subspace.  
In molecule calculations, MO coefficients $C_{\mu p}$ are usually real, and we can assume here that $U_{pq}^x$ is a real matrix. 
Take the first order derivative of Eq. \ref{RHF_SCF} w.r.t. $x$, and one has 
\begin{equation}
\begin{aligned}
    \frac{\partial \epsilon_p}{\partial x}\delta_{pq} & = \frac{\partial F_{pq}}{\partial x} \\
    & = F_{pq}^x + (\epsilon_p-\epsilon_q)U_{pq}^x - \epsilon_qS_{pq}^x + \sum_{mo}{U_{mo}^xA_{pq,om}}
\end{aligned}
\label{RHF_CPHF_1}
\end{equation}
where
\begin{equation}
\begin{aligned}
    F_{pq}^x & = h_{pq}^x + \sum_r{f_r[2(pq|rr)^x-(pr|rq)^x]} \\
    A_{pq,rm} & = 2(pq|rm)+2(pq|mr)-(pm|rq)-(pr|mq)
\end{aligned}
\end{equation}
For $p\neq q$, Eq. \ref{RHF_CPHF_1} equals zero. Using Eq. \ref{Upq_Spqx} and real $U^x$ condition, we have
\begin{equation}
\begin{aligned}
    0 & = F_{pq}^x + (\epsilon_p-\epsilon_q)U_{pq}^x - \epsilon_qS_{pq}^x + \sum_{o'o''}{U_{o''o'}^xA_{pq,o'o''}} + \sum_{o'u'}{U_{u'o'}^xA_{pq,o'u'}} \\
    & = F_{pq}^x + (\epsilon_p-\epsilon_q)U_{pq}^x - \epsilon_qS_{pq}^x - \frac{1}{2}\sum_{o'o''}{S_{o'o''}^xA_{pq,o'o''}} + \sum_{o'u'}{U_{u'o'}^xA_{pq,u'o'}} \\
    & \Rightarrow (\epsilon_q-\epsilon_p)U_{pq}^x - \sum_{u'o'}{U_{u'o'}^xA_{pq,u'o'}} = B_{0,pq}^x
\end{aligned}
\label{RHF_CPHF_2}
\end{equation}
where
\begin{equation}
    B_{0,pq}^x = F_{pq}^x -\epsilon_q S_{pq}^x - \frac{1}{2}\sum_{o'o''}{S_{o'o''}^xA_{pq,o'o''}}
\end{equation}

Eq. \ref{RHF_CPHF_2} is a set of linear equations for $U_{pq}^x$, and is exactly the 1st-order CPHF equation for RHF orbitals. 
Among these equations, those with $(p,q)\in(u,o)$ cannot be decoupled from each other. 
These $n_u\times n_o$ equations should be solved simultaneously.
After $U_{uo}$ is solved, those with $(p,q)\notin(u,o)$ can be directly calculated with the following equation. 
\begin{equation}
    U_{pq}^x = \frac{B_{0,pq}^x + \sum_{u'o'}{U_{u'o'}^xA_{pq,u'o'}}}{\epsilon_q-\epsilon_p}
    \label{RHF_U_nonind}
\end{equation}
If $\epsilon_p=\epsilon_q$ in Eq. \ref{RHF_U_nonind}, $U_{pq}^x$ diverges. 
Fortunately, in our case, we use Eq. \ref{RHF_U_nonind} only for $(p,q)=(b,c) \text{ or } (v,d)$. 
Therefore, the divergence issue does not occur as long as degenerate orbitals are not partially frozen and partially active in CASCI. 

Now let's substitute $U_{pq}^x$ back into Eq. \ref{force_4}, and focus on the last two terms. 
\begin{equation}
\begin{aligned}
    & - \frac{1}{2}\sum_{pq}{(I_{pq}+I_{qp}^*)S_{pq}^x} + \frac{1}{2}\sum_{(p,q)\notin\text{gray}}{(I_{pq}-I_{qp}^*)(U_{pq}^x-U_{qp}^{x*})} \\
    & = -\sum_{(p,q)\in\text{gray}}{I_{pq}S_{pq}^x} + [(\sum_{p>q,(p,q)\notin\text{gray}}{U_{pq}^x(I_{pq}-I_{qp}^*)} - \sum_{p<q,(p,q)\notin\text{gray}}{I_{pq}S_{pq}^x}) + (\text{c.c.})]
\end{aligned}
\label{force_4_term45}
\end{equation}
\begin{equation}
\begin{aligned}
    & \sum_{p>q, (p,q)\notin\text{gray}}{U_{pq}^x(I_{pq}-I_{qp}^*)} \\
    = & \sum_{(p,q)\in\text{slashed}}{U_{pq}^x(I_{pq}-I_{qp}^*)} + \sum_{(p,q)\in(b,c)\text{ or }(v,d)}{U_{pq}^x(I_{pq}-I_{qp}^*)} \\
    = & \sum_{(p,q)\in\text{slashed}}{U_{pq}^x(I_{pq}-I_{qp}^*)} + \sum_{(p,q)\in(b,c)\text{ or }(v,d)}{\frac{I_{pq}-I_{qp}^*}{\epsilon_q-\epsilon_p}(B_{0,pq}^x + \sum_{(m,r)\in\text{slashed}}{U_{mr}^xA_{pq,mr}})} \\
    = & \sum_{(p,q)\in\text{slashed}}{U_{pq}^x H_{pq}} + \sum_{(p,q)\in(b,c)\text{ or }(v,d)}{\frac{I_{pq}-I_{qp}^*}{\epsilon_q-\epsilon_p}B_{0,pq}^x} \\
\end{aligned}
\label{force_4_term45_2}
\end{equation}
where
\begin{equation}
    H_{pq} = I_{pq} - I_{qp}^* + \sum_{(m,r)\in(b,c)\text{ or }(v,d)}{A_{mr,pq}\frac{I_{mr}-I_{rm}^*}{\epsilon_r-\epsilon_m}}
\end{equation}
and ``slashed'' denotes the slashed area in Fig. \ref{orb_partition_SI}, i.e. the $(u,o)$ area.

One can use Z-vector method to simplify the calculation \cite{Yamaguchi_1994}. 
Let's define $V_{pq}$ on $(p,q)\in(u,o)$ such that
\begin{equation}
    (\epsilon_q-\epsilon_p)V_{pq} - \sum_{(m,r)\in\text{slashed}}{A_{mr,pq}V_{mr}} = H_{pq}
\end{equation}
Then one can prove that $\sum_{(p,q)\in\text{slashed}}{V_{pq}B_{0,pq}^x} = \sum_{(p,q)\in\text{slashed}}{U_{pq}^xH_{pq}}$. 
Substitute this into Eq. \ref{force_4_term45_2}, and one has
\begin{equation}
\begin{aligned}
    & \sum_{p>q, (p,q)\notin\text{gray}}{U_{pq}^x(I_{pq}-I_{qp}^*)} \\
    = & \sum_{(p,q)\in\text{slashed}}{V_{pq}B_{0,pq}^x} + \sum_{(p,q)\in(b,c)\text{ or }(v,d)} {\frac{I_{pq}-I_{qp}^*}{\epsilon_q-\epsilon_p}B_{0,pq}^x}
\end{aligned}
\label{force_4_term45_3}
\end{equation}
\begin{equation}
\begin{aligned}
    -F_x = \frac{\partial E_\text{tot}}{\partial x} & = \frac{\partial E_\text{nuc}}{\partial x} + \sum_{pq}{\gamma_{pq}h_{pq}^x} + \frac{1}{2}\sum_{pqrs}{\gamma_{pqrm}(pq|rm)^x} -\sum_{(p,q)\in\text{gray}}{I_{pq}S_{pq}^x} \\
    & + [(\sum_{(p,q)\in\text{slashed}}{V_{pq}B_{0,pq}^x} + \sum_{(p,q)\in(b,c)\text{ or }(v,d)}{\frac{I_{pq}-I_{qp}^*}{\epsilon_q-\epsilon_p}B_{0,pq}^x} - \sum_{p<q,(p,q)\notin\text{gray}}{I_{pq}S_{pq}^x}) \\
    & + (\text{c.c.})]
\end{aligned}
\label{force_4_rhf}
\end{equation}

In Eq. \ref{force_4}, we have to solve $U^x$ for each nuclear degree of freedom $x$. 
In Eq. \ref{force_4_rhf}, we need to solve the CPHF-like equation only once. 
In the work, nuclear forces are calculated with Eq. \ref{force_4_rhf}.

\subsubsection{Formula for ROHF orbitals}

\begin{figure}
    \centering
    \includegraphics[width=8.0cm]{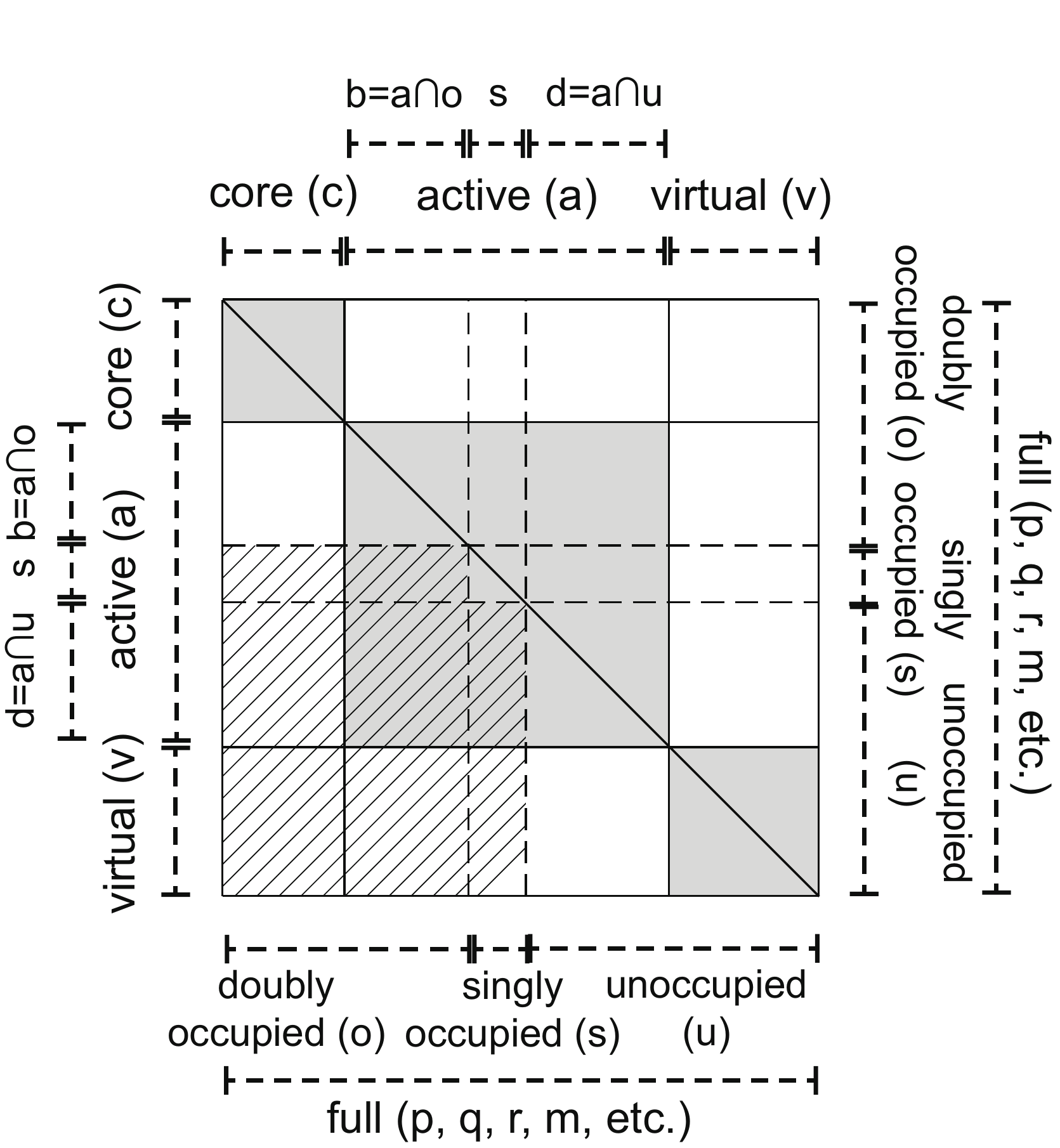}
    \caption{
    The partition of 1-electron ROHF orbitals to construct the Lagrangian matrix $I_{pq}$. 
    From the view of ROHF, the orbitals are classified into three parts, doubly occupied (o), singly occupied (s) and unoccupied ones (u), divided by dashed lines. 
    From the view of CASCI, the orbitals are classified into three parts,  
    core (c, orbitals that are always fully occupied), 
    active (a, orbitals on which FCI expansion is performed), 
    and virtual (v, orbitals that are never occupied) ones, 
    divided by solid lines. 
    The gray area, i.e. $(c,c), (a,a)$ and $(v,v)$, denotes the matrix blocks where the Lagrangian matrix $I_{pq}$ is hermitian, and in which the orbital rotations do not change the wavefunction.  
    The slashed area, i.e. $(u,o), (u,s)$ and $(s,o)$, denotes the matrix blocks where independent $U_{pq}^x$ elements are solved with CPHF equations. }
    \label{ROHF_orb_part}
\end{figure}

The ROHF SCF equation can be formulated as 
\begin{equation}
    F_{pq} = h_{pq} + \sum_r{[2(pq|rr)f_r - (pr|rq)(f_r+\kappa_{pqr})]} = \epsilon_p\delta_{pq}
    \label{ROHF_SCF}
\end{equation}
where 
\begin{equation}
    f_r = \left\{
    \begin{array}{cc}
    1 & r=o, \\
    \frac{1}{2} & r=s, \\
    0 & r=u. 
    \end{array} \right.
    , \quad
    \kappa_{pqr} = \begin{cases}
        0, & r=o \text{ or } r=u \\
        \left(
        \begin{array}{ccc}
            0 & -\frac{1}{2} & 0 \\
            -\frac{1}{2} & 0 & \frac{1}{2} \\
            0 & \frac{1}{2} & 0 
        \end{array}
        \right), & r=s
    \end{cases}
\end{equation}
Here $o, s$ and $u$ denote the ROHF orbital partition (see Fig. \ref{ROHF_orb_part}). 
If $s=\varnothing$, then ROHF equation is reduced to RHF equation. 

Take the first order derivative of Eq. \ref{ROHF_SCF}, we have
\begin{equation}
\begin{aligned}
    \frac{\partial \epsilon_p}{\partial x} & = F_{pq}^x + (\epsilon_p-\epsilon_q)U_{pq}^x - \epsilon_qS_{pq}^x \\
    & + \sum_{ms}{(ps|sm)(\kappa_{pms}-\kappa_{pqs})U_{mq}^x} + \sum_{ms}{(ms|sq)(\kappa_{qms}-\kappa_{qps})U_{mp}^x} \\
    & + \sum_{mr}{U_{mr}^x\tau_{pq,rm}}
\end{aligned}
\label{ROHF_SCF_1}
\end{equation}
where
\begin{equation}
\begin{aligned}
    F_{pq}^x & = h_{pq}^x + \sum_r{[2f_r(pq|rr)^x - (pr|rq)^x(f_r+\kappa_{pqr})]} \\
    \tau_{pq,rm} & = 2f_r[(pq|rm) + (pq|mr)] - (f_r+\kappa_{pqr})[(pm|rq)+(pr|mq)]
\end{aligned}
\end{equation}
When $p\neq q$, Eq. \ref{ROHF_SCF_1} becomes a set of linear equations of $U_{pq}^x$. 

Now we further analyze Eq. \ref{ROHF_SCF_1} and decouple all possible equations from it. 
\begin{equation}
\begin{aligned}
    \sum_{mr}{U_{mr}^x\tau_{pq,rm}} & = \sum_{MR}{\sum_{m\in M,r\in R}{U_{mr}^x\tau_{pq,rm}}} = \sum_R{\sum_{(r_1,r_2)\in R}{U_{r_1r_2}^x\tau_{pq,r_2r_1}}} + \sum_{M\neq R}{\sum_{m\in M,r\in R}{U_{mr}^x\tau_{pq,rm}}} \\
    & = \sum_{M>R}{\sum_{m\in M, r\in R}{U_{mr}^x(\tau_{pq,rm}-\tau_{pq,mr})}}
    - \sum_{M>R}{\sum_{m\in M, r\in R}{S_{rm}^x\tau_{pq,mr}}} \\
    & -\frac{1}{2}\sum_R{\sum_{(r_1,r_2)\in R}{S_{r_1r_2}^x\tau_{pq,r_1r_2}}}
\end{aligned}
\end{equation}
Here capital letters ($M,R$) denote partitions ($o, s$ and $u$). 
The ordering of partitions is chosen as $o<s<u$. 
\begin{equation}
\begin{aligned}
    & \sum_{ms}{(ps|sm)(\kappa_{pms}-\kappa_{pqs})U_{mq}^x} + \sum_{ms}{(ms|sq)(\kappa_{qms}-\kappa_{qps})U_{mp}^x} \\
    = &  \sum_{m\in M>Q, s}{(ps|sm)(\kappa_{pms}-\kappa_{pqs})U_{mq}^x} - \sum_{m\in M<Q, s}{(ps|sm)(\kappa_{pms}-\kappa_{pqs})(U_{qm}^x+S_{qm}^x)} \\
    + & \sum_{m\in M>P, s}{(ms|sq)(\kappa_{qms}-\kappa_{qps})U_{mp}^x} - \sum_{m\in M<P, s}{(ms|sq)(\kappa_{qms}-\kappa_{qps})(U_{pm}^x+S_{pm}^x)} \\
    = & \sum_{M>R}{\sum_{m\in M, r\in R}{[(\kappa_{pms}-\kappa_{prs})(\delta_{rq}(ps|sm)+\delta_{mq}(ps|sr))}} \\
    & {{+ (\kappa_{qms}-\kappa_{qrs})(\delta_{rp}(ms|sq)+\delta_{mp}(rs|sq))]U_{mr}^x}} \\
    & - \sum_{M>R}{\sum_{m\in M, r\in R}{[(\kappa_{pms}-\kappa_{prs})(ps|sr)\delta_{qm} + (\kappa_{qms}-\kappa_{qrs})\delta_{pm}(rs|sq)]S_{rm}^x}}
\end{aligned}
\end{equation}
In conclusion, for $p\neq q$, Eq. \ref{ROHF_SCF_1} can be written as the following shorter form, which is the ROHF-CPHF equation. 
\begin{equation}
    (\epsilon_q-\epsilon_p)U_{pq}^x - \sum_{m\in M, r\in R, M>R}{U_{mr}^x A^\text{ROHF}_{pq,mr}} = B^{x,\text{ROHF}}_{0,pq}
\end{equation}
where
\begin{equation}
\begin{aligned}
    A^\text{ROHF}_{pq,mr} & = \tau_{pq,rm}-\tau_{pq,mr} + \sum_s{(\kappa_{pms}-\kappa_{prs})(\delta_{rq}(ps|sm)+\delta_{mq}(ps|sr))} \\
    & + \sum_s{(\kappa_{qms}-\kappa_{qrs})(\delta_{rp}(ms|sq)+\delta_{mp}(rs|sq))}
\end{aligned}
\end{equation}
\begin{equation}
\begin{aligned}
    B^{x,\text{ROHF}}_{0,pq} & = F_{pq}^x - \epsilon_qS_{pq}^x -\frac{1}{2}\sum_R{\sum_{(r_1,r_2)\in R}{S_{r_1r_2}^x\tau_{pq,r_1r_2}}} \\
    & - \sum_{M>R}{\sum_{m\in M, r\in R}{[\tau_{pq,mr} + (\kappa_{pms}-\kappa_{prs})(ps|sr)\delta_{qm} + (\kappa_{qms}-\kappa_{qrs})\delta_{pm}(rs|sq)]S_{rm}^x}} \\
\end{aligned}
\end{equation}

Now substitute the solution of ROHF-CPHF equation into the energy gradient formula.
Similar to RHF case, we have
\begin{equation}
\begin{aligned}
    -F_x = \frac{\partial E_\text{tot}}{\partial x} & = \frac{\partial E_\text{nuc}}{\partial x} + \sum_{pq}{\gamma_{pq}h_{pq}^x} + \frac{1}{2}\sum_{pqrm}{\gamma_{pqrm}(pq|rm)^x} -\sum_{(p,q)\in\text{gray}}{I_{pq}S_{pq}^x} \\
    & + [(\sum_{(p,q)\in\text{slashed}}{V_{pq}^\text{ROHF} B_{0,pq}^{x,\text{ROHF}}} + \sum_{(p,q)\in(b,c)\text{ or }(v,d)}{\frac{I_{pq}-I_{qp}^*}{\epsilon_q-\epsilon_p}B_{0,pq}^{x,\text{ROHF}}} \\
    & - \sum_{p<q,(p,q)\notin\text{gray}}{I_{pq}S_{pq}^x}) + (\text{c.c.})]
\end{aligned}
\label{force_4_rohf}
\end{equation}
Comparing Eq. \ref{force_4_rohf} with Eq. \ref{force_4_rhf}, one only replaces $B_{0,pq}^x$ with $B_{0,pq}^{x,\text{ROHF}}$, and replaces $V_{pq}$ with $V_{pq}^\text{ROHF}$. 
Here $V_{pq}^\text{ROHF}$ is defined in the slashed block in Fig. \ref{ROHF_orb_part}. 
\begin{equation}
\begin{aligned}
    (\epsilon_q-\epsilon_p)V_{pq}^\text{ROHF} - \sum_{(m,r)\in\text{slashed}}{A_{mr,pq}^\text{ROHF}V_{mr}^\text{ROHF}} = H_{pq}^\text{ROHF} \\
    H_{pq}^\text{ROHF} = I_{pq} - I_{qp}^* + \sum_{(m,r)\in(b,c)\text{or}(v,d)}{A_{mr,pq}^\text{ROHF} \frac{I_{mr}-I_{rm}^*}{\epsilon_r-\epsilon_m}}
\end{aligned}
\end{equation}

\subsection{Conical crossing between $\tilde{X}$ and $\tilde{B}$}

The conical crossing in water monomer has been discussed extensively in literature \cite{Dixon_water_science_1999,Chang_water_nature_2021}. 
In this work, though we mainly focus on the ground state $\tilde{X}$, the conical crossing between $\tilde{X}$ and the ow excited state $\tilde{B}$ is discussed as follows. 

\begin{figure}
    \centering
    \includegraphics[width=16.0cm]{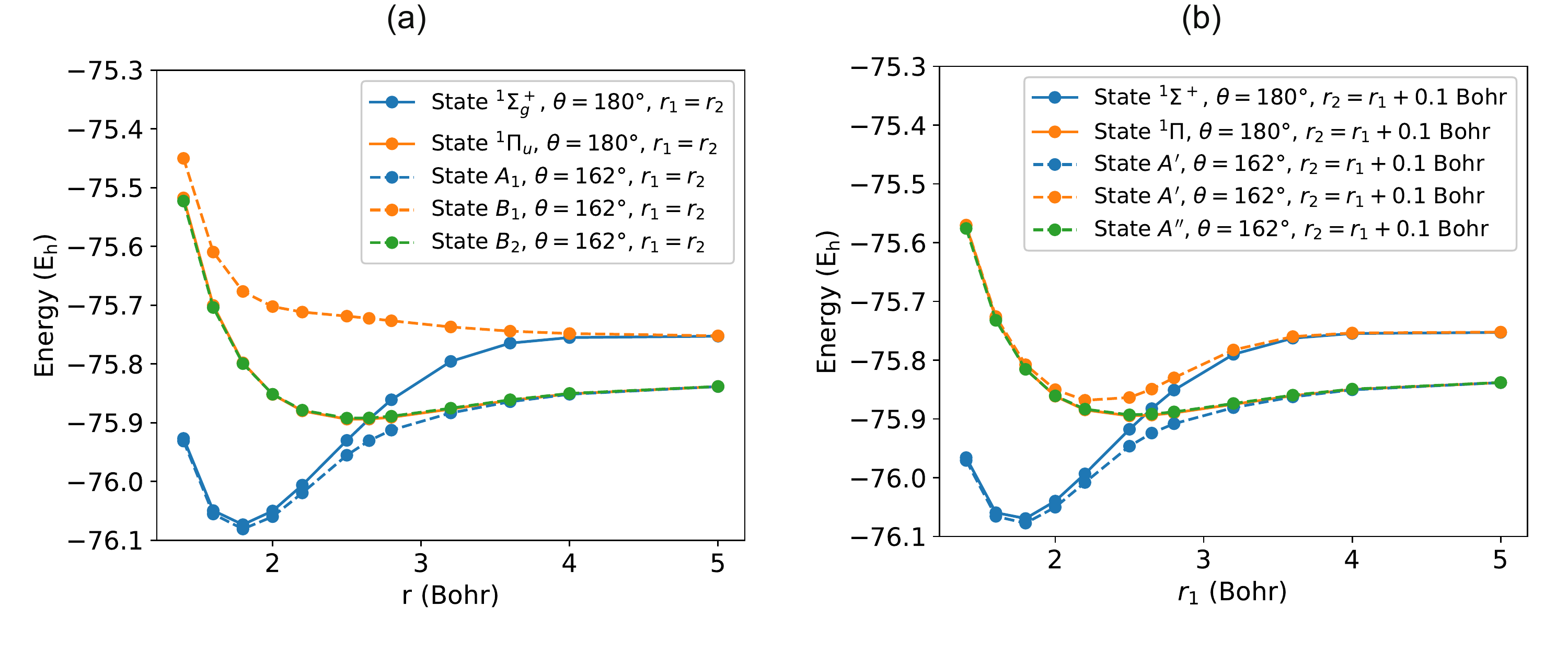}
    \caption{Exact CASCI calculation on conical crossing between $\tilde{X}$ and $\tilde{B}$ near linear and equal-bond-length geometry.
    (a) $r_1=r_2$, $\theta=180\degree$ ($D_{\infty h}$) or $162\degree$ ($C_{2v}$). 
    In $\theta=180\degree$ case, the lowest-energy state of $^1\Sigma_g^+$ and $^1\Pi_u$ are plotted. 
    In $\theta=162\degree$ case, the lowest-energy state of $A_1$, $B_1$ and $B_2$ are plotted. 
    (b) $r_1=r_2 + 0.1\text{ Bohr}$, $\theta=180\degree$ ($C_{\infty v}$) or $162\degree$ ($C_s$). 
    In $\theta=180\degree$ case, the lowest-energy states of $^1\Sigma^+$ and $^1\Pi$ are plotted. 
    In $\theta=162\degree$ case, the 2 lowest-energy states of $A'$, and 1 lowest-energy state of $A''$ are plotted. 
    }
    \label{cc_1}
\end{figure}

Exact FCI diagonalization are performed on H$_2$O monomer with the 6-31G basis set and 1s core of oxygen frozen (CAS(12o,8e)) (Fig. \ref{cc_1}).
A number of structures are selected near the linear ($\theta=180\degree$) and equal-bond-length ($r_1=r_2$) geometry. 
There are 3 low energy states involved in this conical crossing, $\tilde{X} A'$, $\tilde{A} A''$ and $\tilde{B} A'$. 

In Fig. \ref{cc_1}a, $r_1=r_2$ holds. 
At $\theta=162\degree$, the monomer has $C_{2v}$ point group symmetry, and $\tilde{X} A'$, $\tilde{A} A''$ and $\tilde{B} A'$ carry irreducible representations (irrep) $A_1$, $B_2$ and $B_1$ of $C_{2v}$, respectively.
Their energies are shown in blue, green and orange dashed curves, respectively. 
At $\theta=180\degree$, the point group of H$_2$O is raised to $D_{\infty h}$, and the irreps of the 3 states become $^1\Sigma_g^+$ and $^1\Pi_u$, with the latter 2-fold degenerate. 
The $^1\Sigma_g^+$ and $^1\Pi_u$ states of $\theta=180\degree$ and $r_1=r_2$ are shown in Fig. \ref{cc_1}a with blue and orange solid lines, respectively.
A crossing between the two states appears at around 2.7 Bohr.
Their electronic states of $^1\Sigma_g^+$ and $^1\Pi_u$ are adiabatically connected with low symmetry non-linear geometries. 
The way they connect with non-linear geometries in small bond length is different from that in large bond length. 
In geometries with a smaller bond length than the crossing ($r_1<2.7\text{ Bohr}$), $^1\Sigma_g^+$ is connected with $A_1$ ($\tilde{X}$), and the two-fold degenerate $^1\Pi_u$ is connected with $B_1$ ($\tilde{B}$) and $B_2$ ($\tilde{A}$). 
In a larger bond length, $^1\Sigma_g^+$ is connected with $B_1$ ($\tilde{B}$), and $^1\Pi_u$ is connected with $B_2$ ($\tilde{A}$) and $A_1$ ($\tilde{X}$).
Therefore, the crossing between $^1\Sigma_g^+$ and $^1\Pi_u$ is the only intersection point between $\tilde{X}$ and $\tilde{B}$ on the $r_1=r_2$ section.

In $r_1\neq r_2$ case (Fig. \ref{cc_1}b), similar analysis is still valid, if one replaces $D_{\infty h}$ with $C_{\infty v}$, $C_{2v}$ with $C_s$, $^1\Sigma_g^+$ with $^1\Sigma^+$, $^1\Pi_u$ with $^1\Pi$, $B_2$ with $A''$, and $A_1$ and $B_1$ with the two lowest energy $A'$ states respectively. 
In this way, we can also identify the crossing between $^1\Sigma^+$ and $^1\Pi$ as the conical crossing. 

\subsection{Selection routine of structure points in the training set of PES}

The 288 structure points of H$_2$O monomer used as the training set are selected by the following two steps. 
\begin{enumerate}
    \item [(1)] {
    For each 2d section of constant $r_1-r_2$, take a coarse-grained 2d grid, put all grid points into the training set, and calculate their energies and nuclear forces. }
    \item[(2)] {
    Fill some new data points into empty space that comes from invalid data points due to numerical non-convergence. 
    }
    \item[(3)] {
    Perform GPR fit, and add data points near abnormal extreme points or non-smooth points to verify these features. Repeat this step until all features are confirmed or falsified. 
    }
\end{enumerate}


\begin{acknowledgments}
The authors thank Qiming Sun for helpful discussions.
This work was supported by the National Natural Science Foundation of China under Grant No. 11974024 and No. 92165101, the National Key R\&D Program of China under Grant No. 2021YFA1400500, and the
Strategic Priority Research Program of Chinese Academy of Sciences under Grant No. XDB33000000.
We are grateful for computational resources provided by Peking University, the TianHe-1A supercomputer, Shanghai Supercomputer Center, and Songshan Lake Materials Lab.
\end{acknowledgments}

\bibliography{ref.bib}

\end{document}